\documentclass[twocolumn,prl,superscriptaddress,unsortedaddress,showpacs,preprintnumbers,amsmath,amssymb]{revtex4}
\usepackage[dvips]{graphicx}
\usepackage{dcolumn}
\usepackage{amsmath}
\usepackage{times}

\begin{document}


\title{Positron annihilation in $e^+ -$H collision above the Positronium
formation threshold}

\author{S.L. Yakovlev}
\affiliation{
Department of Computational Physics, St Petersburg State University,
St Petersburg, Russia 198504
}
\author{ C-.Y. Hu}
\affiliation{
Department of Physics and Astronomy,
California State University, Long Beach, California 90840
}

\date{\today}

\begin{abstract}
A long-standing problem to account for the electron-positron
annihilation in positron Hydrogen scattering above the Positronium formation
threshold
has been resolved by the use of the  three-body Faddeev
formalism.
The multichannel three-body theory for scattering states in presence of
a complex absorbing potential
is developed in order
to compute the direct $e^+ e^-$ annihilation amplitude, the amplitude of
Positronium formation and
respective cross sections.
A number of $e^+ e^-$ direct annihilation cross sections and Positronium
formation
cross sections in the energy gap between Ps$(1s)$ and H$(n=2)$ thresholds
are
reported for both the positron-Hydrogen incoming channel as well as the
proton-Positronium incoming channel.
\end{abstract}
\maketitle
When colliding with the Hydrogen atom in the energy region between
Positronium Ps$(1s)$ and
Hydrogen H$(n=2)$ thresholds,
the positron can annihilate in two possible channels either in the
elastic $e^{+}-$H one
or in the rearrangement $p-$Ps one.
During rearrangement the positron captures
 the electron from H and forms Positronium in the Ps$(1s)$ state and
 then the positron-electron pair
annihilates from this state.
The lifetime of Ps$(1s)$ is known to
depend on the total spin \cite{rich}.
This well known annihilation mechanism is of the two-body nature,
{\it i.e.},
once Positronium is formed the third particle does not affect the
annihilation process. The annihilation
mechanism in the elastic $e^+-$H channel is more complicated and needs
three-body footing.
There was a number of attempts for the unified three-body treatment of
the positron annihilation phenomenon
and the Positronium formation,
especially in the vicinity of the Positronium formation threshold
\cite{mitroy}, \cite{iks}, \cite{gl}.
Nevertheless as to the best of our knowledge, the self-consistent
multichannel theory for the annihilation
in positron-atom collisions when
the rearrangement channel is open does not exist to 
date. It is the aim with this paper
to fill out this gap by combining the three-body multichannel scattering
approach basing on
modified Faddeev equations (MFE) \cite{fadd}, \cite{merkuriev} with the
concept of a complex
absorbing potential \cite{mitroy} which
allows to treat the annihilation in the framework of non-relativistic
quantum mechanics.
In the paper we restrict ourselves within the Ore gap energy region.
The extension onto the case when more than one Hydrogen
(and Positronium) state are participating is straightforward.

Let us recapitulate those portions of the one channel approach to the
description
of the annihilation phenomenon in
positron Hydrogen scattering below the rearrangement threshold which are
essential
for a multichannel extension. The well known QED formula \cite{fraser},
\cite{charlton}
\begin{equation}
{\overline \sigma}^{a}=\pi r^2_0 (c/v){\overline Z}_{\mbox{eff}}
\label{QEDsigma}
\end{equation}
expresses the spin averaged annihilation cross section
${\overline {\sigma}^a}=
{\small 1/4}\, {^1}\sigma^a + 
{\small 3/4}\, {^3}\sigma^a $
in terms of the effective number of electrons ${\overline Z}_{\mbox{eff}}$
contributing to the annihilation.
The respective number ${^{2S+1}}Z_{\mbox{eff}}$ for a spin state $S$
is given by the integral
\begin{equation}
{^{2S+1}}Z_{\mbox{eff}}= \int d{\bf r}_1 d{\bf r}_3\,
|{^{2S+1}}\Psi^0({\bf r}_1,{\bf r}_3)|^{2}
\delta({\bf r}_1-{\bf r}_3).
\label{Zeff}
\end{equation}
Throughout the paper we use bold letters for vectors, {\it e.g.},
${\bf r}$, not bold ones for
their magnitudes, {\it e.g.}, $r=|{\bf r}|$, and we choose units such that
$\hbar =1$.
In formulae (\ref{QEDsigma}) and (\ref{Zeff}) the ingredients has the
following meaning, $r_0$ is the
classical electron radius, $c$ is the velocity of light, $v$ is the
incident velocity of
positron,  ${^{2S+1}}\Psi^0$
is the $e^+-$H scattering wave function when only Coulomb interaction is
taken into account, and
${\bf r}_1$ and ${\bf r}_3$ are position vectors of positron and electron
in
the coordinate frame associated with proton. The presence of the
delta-function in
(\ref{Zeff}) manifests the fact that $e^+$ and $e^-$ annihilate when they
occupy the same position.
The non-relativistic quantum mechanical description consistent with the
above formulation can be based on the
concept of a complex absorbing potential \cite{mitroy}, \cite{iks}. The use
of this potential
in the Schr\"odinger equation leads to the particle
loss. The cross section associated with this loss is actually  the
annihilation cross section. The
absorbing potential is defined by \cite{iks}
\begin{equation}
i g W({\bf r}_1-{\bf r}_3)\equiv -i (e^2/a_0) (^{2S+1}c)
\alpha^{S+3}\delta({\bf r}_1-{\bf r}_3),
\label{W}
\end{equation}
where $e$ is the electron charge, $a_0$ is the Bohr radius, $\alpha$ is
the fine structure constant,
and spin dependent constants ${^{2S+1}}c$ are defined as
$
{^1}c=2\pi \ \ \ \mbox{and}\ \ \ {^3}c=8(\pi^2-9)/9.
$

The Schr\"odinger equation for the positron Hydrogen wave function with
outgoing asymptotic boundary
conditions
\begin{equation*}
(E-H^0)\Psi^+_{{\bf p}_1}=igW\Psi^+_{{\bf p}_1},
\end{equation*}
where $H^0$ is the three-body Coulomb Hamiltonian and ${{\bf p}_1}$ is the
momentum of incident positron,
can be rewritten as
\begin{equation}
\Psi^+_{{\bf p}_1}=\Psi^{0+}_{{\bf p}_1}+ig(E^{+}-H^0)^{-1}
W\Psi^+_{{\bf p}_1}.
\label{Scri}
\end{equation}
Here  $\Psi^{0+}_{{\bf p}_1}$ is the solution to
the pure three-body Coulomb problem
$(H^0-E)\Psi^{0+}_{{\bf p}_1}=0$ and $E^{+}=E+i0$.
It is seen from (\ref{Scri}) that
the asymptotics of $\Psi^+_{{\bf p}_1}$, as $r_1$ approaches infinity,
takes the form
\begin{equation}
\Psi^+_{{\bf p}_1}({\bf r}_1,{\bf r}_3) \sim 
\psi({\bf r}_3)\{ e^{ i{\bf p}_1\cdot{\bf r}_1}
+
\frac{e^{ip_1r_1}}{r_1}
 [F_{e} -igF_a]
 \}.
\label{psi-as}
\end{equation}
In (\ref{psi-as}), $\psi({\bf r}_3)$ is the Hydrogen ground state wave
function.
The elastic scattering amplitude $F_{e}$ is entirely due to the
$\Psi^{0+}_{{\bf p}_1}$ contribution,
and the annihilation amplitude $F_a$ is due to the absorbing potential.
This amplitude is given by
\begin{equation}
F_{a}({\bf p}'_1,{\bf p}_1)=
\frac{m}{2\pi}\langle \Psi^{0-}_{{\bf p}'_1}|W|\Psi^{+}_{{\bf p}_1}
\rangle,
\label{Fa}
\end{equation}
where $m$ is the electron mass.
In principle, the annihilation cross section $\sigma^{a}$ can be computed
from the total cross section $\sigma^{t}$ as
\begin{equation}
\sigma^a=\sigma^{t} - \int d\Omega\, |F_{e}|^2,\ \ \ \sigma^{t}=\int
d\Omega \,|F_{e}-igF_a|^{2},
\label{sigmatot}
\end{equation}
but in the one channel case there is an alternative
way which is based on the optical theorem
\begin{equation*}
\sigma_{t}= \frac{4\pi}{p_1}\Im \mbox{m} [F_{e}({\bf p}_1,{\bf p}_1)
-igF_a({\bf p}_1,{\bf p}_1)],
\end{equation*}
from which the  annihilation cross section appears as
\begin{equation}
\sigma_a=g\frac{4\pi}{p_1}\Im \mbox{m} [-iF_a({\bf p}_1,{\bf p}_1)].
\label{sigma-a-o}
\end{equation}
Now, the annihilation cross section (\ref{QEDsigma}) and the
formula (\ref{Zeff}) for the effective number $Z_{\mbox{eff}}$ follow
immediately from (\ref{Fa})
and (\ref{sigma-a-o}) with replacing $\Psi^+_{{\bf p}_1}$ in (\ref{Fa}) by
$\Psi^{0+}_{{\bf p}_1}$ which,
considering (\ref{Scri}) perturbatively  in view of $|g|\ll 1$,
is the first
order approximation for $\Psi^+_{{\bf p}_1}$, {\it i.e.},
$
\Psi^+_{{\bf p}_1}=\Psi^{0+}_{{\bf p}_1} + O(g)
$.

Let us now turn to the case when the Positronium formation channel is open
and thus the $e^+-$H collision becomes
truly multi arrangemental. In this case the integral defined by (\ref{Zeff})
diverges and the cross section defined by
formula (\ref{QEDsigma}) is infinite. On the contrary, the exact amplitudes
$F_{e}$ and $F_{a}$ defined by
(\ref{psi-as}), (\ref{Fa}) are finite (as long as the exact solution  to (
\ref{Scri}) is assumed)
due to the effect of the
absorbing potential which makes the Positronium binding energy complex.
Although the exact amplitudes are
well defined, the matrix element (\ref{Fa}) cannot remedy the problem with
the annihilation cross section. Indeed,
the contribution of elastic and rearrangement channels to the optical
theorem for the three body
system \cite{schmid} is additive
\begin{equation}
\frac{4\pi}{p_1}\Im \mbox{m} [F_{e}({{\bf p}_1},{\bf p}_1)
-igF_{a}({{\bf p}_1},{\bf p}_1)]=
\sigma_{11}+\sigma_{21} .
\label{3boptic}
\end{equation}
Here $\sigma_{11}$ is the total (elastic plus annihilation) cross section
in the $e^+-$H channel and $\sigma_{21}$
is the total Positronium formation (and annihilation after Positronium
formation) cross section.
Since the contribution of the annihilation in the left hand side of
(\ref{3boptic}) is proportional to the matrix
element (\ref{Fa}) it is apparent  that with this quantity the direct
annihilation process and the process
of the annihilation after Positronium formation are {\it inseparable}.
We believe that a variant of this effect of inseparability was observed
in the numerical study of positron Hydrogen scattering
in \cite{iks}, \cite{igarashi2}.

So, it is clear that the shortcoming of the outlined formalism is
due to an inadequate treatment
of multichannel scattering.
The approach which is proven to define correctly the transition amplitudes
between all open channels
in the three-body case
is the Faddeev equation formalism
\cite{fadd} and its modification MFE
for long-range interactions \cite{merkuriev}.
In this approach the wave function is split into the sum
of components each of them corresponding to the given asymptotic
arrangement. The solution of the set of equations
for components (Faddeev equations) satisfies the following requirements:
 i) the sum of components
obeys the Schr\"odinger equation; ii) the asymptotics of the component
associated with a given
asymptotic arrangement
includes contributions from all open channels in this arrangement and
coincides with the restriction of the
wave function on this arrangement \cite{merkuriev}. This last property
guaranties that the amplitudes
computed from components are exact physical scattering amplitudes.

The three-body relative Jacobi coordinates ${\bf x}_i$ and ${\bf y}_i$
are used to describe the configuration
of the three-body
system in the c.m. frame. So it is not required for proton to be infinitely
heavy.
The index $i$ runs over all three arrangements of the system and
corresponds to the spectator particle $i$. The positron Hydrogen Hamiltonian reads
\begin{equation}
H=T+\frac{e_2e_3}{x_1}+\frac{e_3e_1}{x_2}+\frac{e_1e_2}{x_3}
+ igW_2({\bf x}_2).
\label{ham}
\end{equation}
Here $T$ is the c.m. kinetic energy, the index 1 corresponds to the
positron, whereas 2 and 3 do to proton
and electron, respectively, thus $e_1=e_2=e$, $e_3=-e$. The last term $igW_2$ is the
absorbing potential (\ref{W}).
For long-range potentials the modification \cite{merkuriev}
is needed, which consists in splitting  of the Coulomb potentials into
long-range $V^l_i$ and short-range
$V^s_i$ pieces  such that
\begin{equation}
V^l_i({\bf x}_i,{\bf y}_i)+V^s_i({\bf x}_i,{\bf y}_i)=V^C_i({\bf x}_i)
\equiv \frac{e_je_k}{x_i}.
\label{split}
\end{equation}
This splitting is made in the three-body configuration space by a smooth
splitting function $\zeta_i({\bf x}_i,{\bf y}_i)$
constructed in such a way that $\zeta_i({\bf x}_i,{\bf y}_i)=1$ if
$x_i/x_0<(1+y_i/y_0)^\nu$ and
$\zeta_i({\bf x}_i,{\bf y}_i)=0$ if $x_i/x_0>(1+y_i/y_0)^\nu$ for some
$x_0>0,\, y_0>0$ and $0<\nu<1/2$.
With such a $\zeta_i$ the short- and long-range parts of the Coulomb
potentials are defined as
$$
V^s_i({\bf x}_i,{\bf y}_i)=\zeta_i({\bf x}_i,{\bf y}_i)V^C_i({\bf x}_i);
\ \ \
V^l_i=V^C_i-V^s_i.
$$
The Hamiltonian (\ref{ham}) is transformed then into
\begin{equation*}
H=H^l+V^s_1+V^s_2+igW_2; \ \ \ H^l=T+V^l_1+V^l_2+V^C_3.
\end{equation*}
Faddeev components of the wave function $\Psi^+$ are defined by formulae
\begin{eqnarray*}
\Psi^+_1&=&(E^{+} -H^l)^{-1}V^s_1\Psi^+ \\
\Psi^+_2&=&(E^{+}-H^l)^{-1}(V^s_2+igW_2)\Psi^+ ,
\end{eqnarray*}
where $E^{+}=E+i0$. It is straightforward to see that the sum of the
components gives the wave function
$$
\Psi^+=\Psi^+_1+\Psi^+_2
$$
and the components obey the set of MFE
\begin{eqnarray}
(E -H^l-V^s_1)\Psi^+_1&=&V^s_1\Psi^+_2 \nonumber \\
\label{mfe2}
(E-H^l-V^s_2-igW_2)\Psi^+_2&=&(V^s_2+igW_2)\Psi^+_1.
\end{eqnarray}
The important feature of equations (\ref{mfe2}), with regard to description
of the annihilation,
is the fact that the
two-body annihilation
potential $igW_2$ is incorporated into equations in two manners,
{\it i.e.},
in the diagonal part on the left hand side
of the equation (\ref{mfe2}) being responsible in the channel 2 for
annihilation of Positronium after its
formation, and in the
coupling term on the right hand side supporting the annihilation process
in the positron Hydrogen channel 1 through
the coupling.

The set of MFE (\ref{mfe2})
treats the multichannel nature of the problem above the Positronium
formation threshold correctly. Namely, in the asymptotic part of the
channel 1 where $|{\bf y}_1|\to \infty$ and
${\bf x}_1$ is confined within the Hydrogen atom volume, the only
nonvanishing potentials are $V^l_1$ and $V^s_1$
in the combination $V^l_1+V^s_1=V^C_1$ and the first MFE  takes the form
$$
(E-T-V^C_1)\Psi^+_1=0.
$$
So, it is obvious that asymptotically
$$
\Psi^+_1({\bf x}_1,{\bf y}_1) \sim 
\phi_1({\bf x}_1)e^{i{\bf p}_1\cdot {\bf y}_1},
$$
where $\phi_1({\bf x}_1)$
is the Hydrogen wave function with the energy $\epsilon_1$, so that
$E={\bf p}_1^2/2\nu_1+\epsilon_1$, and
$\nu_1$ is the $e^+$-H reduced mass.
In the asymptotic part of the channel 2, where $|{\bf y}_2|\to \infty$
and ${\bf x}_2$ is confined in the volume
of the Positronium atom the nonvanishing potential is $V^C_2+igW_2$ and
the second MFE  is reduced to
$$
(E-T-V^C_2-igW_2)\Psi^+_2=0.
$$
Hence, the asymptotics of the component $\Psi^+_2$ can be written as
$$
\Psi^+_2({\bf x}_2,{\bf y}_2) \sim 
\phi_2({\bf x}_2)e^{i{\bf p}_2\cdot {\bf y}_2},\ \ \
E={\bf p}_2^2/2\nu_2 +\epsilon_2
$$
where $\nu_2$ is the $p$-Ps reduced mass,
$\phi_2({\bf x}_2)$ is the Positronium wave function in  presence of the
absorbing potential, {\it i.e.},
the solution to the equation
\begin{equation}
[-\frac{\Delta_{{\bf x}_2}}{2m} -\frac{e^2}{x_2} +igW_2({\bf x}_2)]
\phi_2({\bf x}_2)=\epsilon_2\phi_2({\bf x}_2).
\label{C-Positronium}
\end{equation}
Let us notice, that the first order approximation with respect to $g$ gives
for
$\epsilon_2$ the formula
$$
\epsilon_2\sim \epsilon^0_2 +ig\langle \phi^0_2|W_2|\phi^0_2\rangle
$$
with $\epsilon^0_2$ and $\phi^0_2$ defined as the solution to
(\ref{C-Positronium}) if $g=0$, {\it i.e.},
the pure Coulomb problem for Positronium.

The above arguments can be
formalized rigorously
with the use of integral equations.
To this end it is convenient to introduce matrix notations
\begin{equation*}
{\bf H}(g)=\left[
            \begin{array}{cc}
            H^l+V^s_1   &    0\\
            0           &H^l+V^s_2+igW_2
            \end{array}
            \right]
\end{equation*}
\vskip 0.3cm
$$
{\bf V}=\left[
            \begin{array}{cc}
             0     & V^s_1\\
             V^s_2 & 0
             \end{array}
         \right],
\ \
{\bf W}=\left[
            \begin{array}{cc}
             0     & 0\\
             W_2 & 0
             \end{array}
         \right],
\ \
{\bf I}=\left[
            \begin{array}{cc}
             1     & 0\\
             0     & 1
             \end{array}
         \right].
$$
Thus, the MFE set takes the form
\begin{equation}
[E{\bf I}-{\bf H}(g)-{\bf V}]{\bf \Psi}=ig{\bf W}{\bf \Psi},
\label{MatrixMFE}
\end{equation}
where ${\protect \bf \Psi}=(\Psi^+_1,\Psi^+_2)^{T}$. The equation
(\ref{MatrixMFE}) can be rewritten as
\begin{equation}
{\bf \Psi}={\bf \Psi}^0+ig[E^{+}{\bf I}-{\bf H}(g)-{\bf V}]^{-1}
{\bf W}{\bf \Psi}.
\label{MatrixIMFE}
\end{equation}
Here ${\bf \Psi}^0$ is the solution to the homogeneous equation
\begin{equation}
[E{\bf I}-{\bf H}(g)-{\bf V}]{\bf \Psi}^0=0
\label{DMFEg0}
\end{equation}
which can be transformed to the integral form
\begin{equation}
{\bf \Psi}^0={\bf \Phi}+[E^{+}{\bf I}-{\bf H}(g)]^{-1}{\bf V}{\bf \Psi}^0 .
\label{IMFEg0}
\end{equation}
The driving term ${\bf \Phi}$ obeys the homogeneous equation
$$
[E{\bf I}-{\bf H}(g)]{\bf \Phi}=0
$$
and its components ${\bf \Phi}=(\Phi_1,\Phi_2)^{T}$
are given in terms of channel functions $\phi_i({\bf x}_i)$ by formulae
$$
\Phi_i({\bf x}_i,{\bf y}_i)=\phi_{i}({\bf x}_i)e^{i{\bf p}_i\cdot
{\bf y}_i}\delta_{ik}
$$
which define through the index $k=1,2$ of the Kronecker-delta the initial
state of the $e^+-$H system.

From the analysis made in \cite{fadd}, \cite{merkuriev} it follows that
matrix integral equations
(\ref{MatrixIMFE}) and (\ref{IMFEg0}) are well
defined and possess the property that the driving term defines the solution
uniquely.
So that, the asymptotics of the solutions of (\ref{MatrixIMFE}) and
(\ref{IMFEg0}) is
entirely due to the asymptotics of the resolvent operators incorporated
in the kernels of these equations.
The latter asymptotics can be evaluated with the help of techniques
developed
in \cite{fadd}, \cite{merkuriev},
\cite{yakovlev}
which leads to the following result for
the solution to (\ref{MatrixIMFE}) as $y_i \to \infty$
\begin{equation*}
\Psi^{+}_i({\bf x}_i,{\bf y}_i,{\bf p}_k)\sim \phi_{i}({\bf x}_i)
[e^{i{\bf p}_i\cdot {\bf y}_i}\delta_{ik}+
\frac{e^{ip_iy_i}}{y_i}F_{ik}(p_i{\bf {\hat y}}_i,{\bf p}_k)].
\end{equation*}
The amplitudes $F_{ik}$ have the structure which is similar to that of
one channel case (\ref{psi-as})
\begin{equation*}
F_{ik}({\bf p}'_i,{\bf p}_k)=F^0_{ik}({\bf p}'_i,{\bf p}_k)
-igF^1_{ik}({\bf p}'_i,{\bf p}_k).
\end{equation*}
The first term $F^0_{ik}$ is entirely due to the contribution
of ${\bf \Psi^0}$ in (\ref{MatrixIMFE}) and if $i=k$ then
$F^0_{kk}$ is
the elastic scattering amplitude whereas  $F^0_{ik}$  is the rearrangement
amplitude if $i\ne k$.
The second term is due to the absorbing potential.
The  direct annihilation amplitude in the $e^+-$H channel $F^1_{1k}$
is given by the formula
\begin{equation}
F^1_{1k}({\bf p}'_1,{\bf p}_k) =\frac{\nu_1}{2\pi}
\langle 
\Psi^{0-}_1({\bf p}'_1)+\Psi^{0-}_2({\bf p}'_1)
|W_2|\Psi^+_1({\bf p}_k)\rangle .
\label{F1}
\end{equation}
Components $\Psi^{0-}_i({\bf p}'_1)$ are solutions to (\ref{IMFEg0}) where
the energy
should be taken on the lower rim to provide incoming boundary conditions,
so that $E^{+}$ should be replaced by $E^{-}=E-i0$, and the inhomogeneous
term is set as
$\Phi_i({\bf x}_i,{\bf y}_i)=\phi_{i}({\bf x}_i)e^{i{\bf p}_i\cdot
{\bf y}_i}\delta_{i1}$.
The amplitude $F^1_{2k}$ has the structure similar to that of (\ref{F1}).
It plays the role of  the correction term either for the Positronium
formation amplitude $(k=1)$ or
for the elastic amplitude of Positronium-proton scattering $(k=2)$.

Once amplitudes of all processes are determined, the cross sections of
interests are defined by
 integrals
\begin{equation}
\sigma_{1k}=\frac{v_1}{v_k}\int d\Omega \;|F^0_{1k}|^2 ;\ \ \
\sigma_{2k}=\frac{v_2}{v_k}\int d\Omega \;|F^0_{2k}-igF^1_{2k}|^2
\label{sigma-ep}
\end{equation}
\begin{equation}
\sigma^a_{1k}=\frac{v_1}{v_k}\int d\Omega \;  [2g\;\Im\mbox{m}F^0_{1k}
F^1_{1k}+g^2|F^1_{1k}|^2].
\label{sigma-ah}
\end{equation}
Here we have split the total cross section in the channel 1 into two
parts similarly to (\ref{sigmatot}),
so that $\sigma^{t}_{1k}=\sigma_{1k}+\sigma^a_{1k}$, where $\sigma^a_{1k}$
is the annihilation cross section.
Formulae (\ref{sigma-ep}), (\ref{sigma-ah}) were used for calculations of
the direct annihilation
 cross section $\sigma^a_{11}$ and the Positronium formation  cross section
 $\sigma_{21}$
for the
$e^+-$H incoming channel, and
the direct annihilation cross section $\sigma^a_{12}$ and the
Hydrogen formation cross section $\sigma_{12}$ when
$p-$Ps is the incoming channel.
During the calculation we systematically used the
smallness of the coupling constant $g$ of the absorbing potential  to
simplify amplitudes,
namely, from (\ref{MatrixIMFE}) it follows that
${\bf \Psi}={\bf \Psi^0}+O(g)$ and from (\ref{IMFEg0}) it is immediately
seen that
 ${\bf \Psi^0}={\bf \Psi^{00}}+O(g)$,
where ${\bf \Psi^{00}}$ is the solution to (\ref{IMFEg0}) when $g=0$.
Hence,   the amplitudes
and cross sections in the leading order can be calculated with the solution
of MFE for the pure three-body Coulomb problem.

In tables below we list results of our calculations for $\sigma_{12}$,
$\sigma_{21}$ and
$\sigma^a_{11}$, $\sigma^a_{12}$ cross sections for total angular momenta
$L=0$, $L=1$ and $L=2$
for five values of the energy of the spectator  in the Ore gap.
All calculations were performed on the base of a quintic spline
algorithm for the solution of differential MFE equations from \cite{hu}.
The numerical error
for cross sections was estimated as not exceeding 5\% for $\sigma^a_{11}$
and 10 \% for $\sigma^a_{12}$.

As a concluding remark we want to emphasis  that  the MFE formalism
presented here
is exact above the Positronium
formation threshold, at the same time it is equivalent to the standard one
below the Positronium formation
threshold \cite{kwh} and it is also  exact for the Positronium ion Ps$^-$
\cite{kkmy}.

The authors appreciate the support of the NSF grant Phy-0243740 and the
generous supercomputer
time awards from
grants MCA96T011 and TG-MCA96T011 under the NSF partnership for Advanced
Computational Infrastructure,
Distributed Terascale Facility (DTF) to the Extensible Terascale Facility.
In  particular we are thankful to PSC and SDSC.
\begin{table}[]
\caption{ Ps formation cross section $\sigma_{21}$, H formation cross
section $\sigma_{12}$ and
direct annihilation cross sections $\sigma^a_{11}$ and $\sigma^a_{12}$
for $L=0$.
Momenta $k$ are in units of $1/a_0$, and $\sigma$'s are in units of
$\pi a^2_0$.
\label{tab_1}}
\begin{center}
\begin{tabular}{rrrrrr
}\hline
$k$           & 0.8000     & 0.8500     & 0.8612     & 0.8615
&  0.8618    \\ \hline
$\sigma_{21}$ & 0.4965[-2]  & 0.5711[-2]  & 0.3239[-1]  & 0.826[-3]
& 0.3259[-2] \\ 
$\sigma_{12}$ & 0.1137[-1]   & 0.9300[-2]   & 0.4982[-1]   & 0.1287[-2]
& 0.4972[-2] \\ 
$\sigma^a_{11}$& 0.299[-6]  & 0.329[-6]  & 0.399[-6]  & 0.332[-6]
& 0.336[-6]  \\ 
$\sigma^a_{12}$& 0.30[-7]   & 0.37 [-7]  & 0.15 [-6]  & 0.53 [-8]
& 0.17 [-7]   \\ \hline
\end{tabular}
\end{center}
\end{table}
\begin{table}[h]
\caption{Same as in TAB. I for $L=1$ with same units.
\label{tab_2}}
\begin{center}
\begin{tabular}{rrrrrr
}\hline
$k$           & 0.8000     & 0.8500     & 0.86315     & 0.8632
&  0.86325    \\ \hline
$\sigma_{21}$ & 0.4793      & 0.5621      & 0.1022[+1]   & 0.6847[-1]
& 0.4077     \\ 
$\sigma_{12}$ & 0.3633       & 0.3032      & 0.5167        & 0.3459  [-1]
& 0.2059      \\ 
$\sigma^a_{11}$& 0.581[-6]  & 0.403[-6]  & 0.574[-6]  & 0.236[-6]
& 0.477[-6]  \\ 
$\sigma^a_{12}$& 0.64[-6]  & 0.12[-5]  & 0.73[-6]  & 0.64[-7]  & 0.12[-6]
\\ \hline
\end{tabular}
\end{center}
\end{table}
\begin{table}[]
\caption{Same as in TAB 1 for $L=2$ with same units.
\label{tab_3}}
\begin{center}
\begin{tabular}{rrrrrr
}\hline
$k$           & 0.8000     & 0.8500     & 0.86549    &0.865505
& 0.86552    \\ \hline
$\sigma_{21}$ & 0.8711     & 0.1167[+1] & 0.8086     &0.1788[+1]
& 0.1306[+1] \\ 
$\sigma_{12}$ & 0.3961     & 0.3779     & 0.2425     &0.5362
& 0.3917     \\ 
$\sigma^a_{11}$& 0.283[-6] & 0.271[-6]  & 0.391[-6] &0.271[-6]
& 0.205[-6]  \\ 
$\sigma^a_{12}$& 0.99[-6]  & 0.12[-5]  & 0.11[-6]&0.19[-5] & 0.58[-6]  \\
\hline
\end{tabular}
\end{center}
\end{table}


\end{document}